\documentclass[11pt,twoside]{article}
\usepackage{./asp2014}
\usepackage{natbib}
\aspSuppressVolSlug
\resetcounters
\usepackage{ wasysym }

\usepackage{url}
\begin{document}

\title{Formation and evolution of exoplanets in different environments}

\author{Vardan~Adibekyan$^1$
\affil{$^1$Instituto de Astrof\'isica e Ci\^encias do Espa\c{c}o, Universidade do Porto, CAUP, Rua das Estrelas, 4150-762 Porto, Portugal; \email{vadibekyan@astro.up.pt}}}


\begin{abstract}

The ultimate goal of exoplanetologists is to discover life outside our Earth and to fully understand our place in
the Universe. Even though we have never been closer to attaining this goal, we still need to understand how and
where the planets (efficiently) form. In this manuscript I briefly discuss the important role of stellar metallicity and chemistry 
on the formation and evolution of exoplanets.

\end{abstract}

\vspace{-0.9cm}

\section{Introduction}

Ever since the first giant exoplanet was discovered orbiting a Sun-like star about twenty years ago \citep{Mayor-95},
the search has been ongoing for small, rocky planets around other stars, evocative of Earth and other
terrestrial planets in the Solar System. As of today, there are more than 3500 planets detected\footnote{exoplanet.eu} and
several thousand candidates \citep{Coughlin-16} waiting for validation. These discoveries helped us to understand that extra-solar planets are
very common in our Galaxy. The diversity of the discovered planets is astonishing, and most of detected planets
brought us more questions than answers. While the Universe is full of surprises, we (exoplanetologists) are
drawing closer to the answer to the most daring questions of humankind: are we alone in the Universe and what
is our place in there? In fact, the last ``simplistic'' calculations\footnote{The authors did not consider requirements of
individual elemental abundances for planet formation \citep[e.g.][]{Adibekyan-12b, Adibekyan-15}.} of \citet{Behroozi-15}
shows that the chance that we are the only civilization the Universe will ever have is less than 8\%.

In this manuscript first I start by briefly presenting how different are the properties of so-far detected exoplanets and how the two completely (ideologically)
different theories are getting close to explain the formation and evolution of these planets. In the second part of the paper I discuss the importance 
of chemical conditions of the environment where the exoplanets form.

\vspace{-0.5cm}

\section{The zoo of exoplanets and their formation scenarios}

Among the few thousands of the detected exoplanets, we observed many words that are very different from what we have in our Solar System and from what we could imagine.
We observed a planet  with an extremely eccentric orbit of 0.97 \citep[HD20782b --][]{O'Toole-09}, a planet in a circumbinary orbit 
surrounded by four suns \citep[PH1b\footnote{This planet was first discovered by two citizen scientists.}/Kepler-64b --][]{Schwamb-13},
a dense ``superplanet''\footnote{Note that this sub-stellar object can be a low-mass brown-dwarf.} with a radius of 
only $\sim$ 1R$_{\jupiter}$ and mass of $\sim$ 22M$_{\jupiter}$ \citep[CoRoT-3b --][]{Deleuil-08}, a very hot planet with a surface day-side temperature of more than 
9000 K orbiting its pulsating hot sun in less than six hours \citep[Kepler-70b/KOI-55.01 --][]{Charpinet-11}.
The detection of these weird exoplanets\footnote{\url{https://en.wikipedia.org/wiki/List_of_exoplanet_extremes}} 
puts to shame many old science fiction stories and make harder the job of new science fiction writers.

\articlefigure[width=0.9\textwidth]{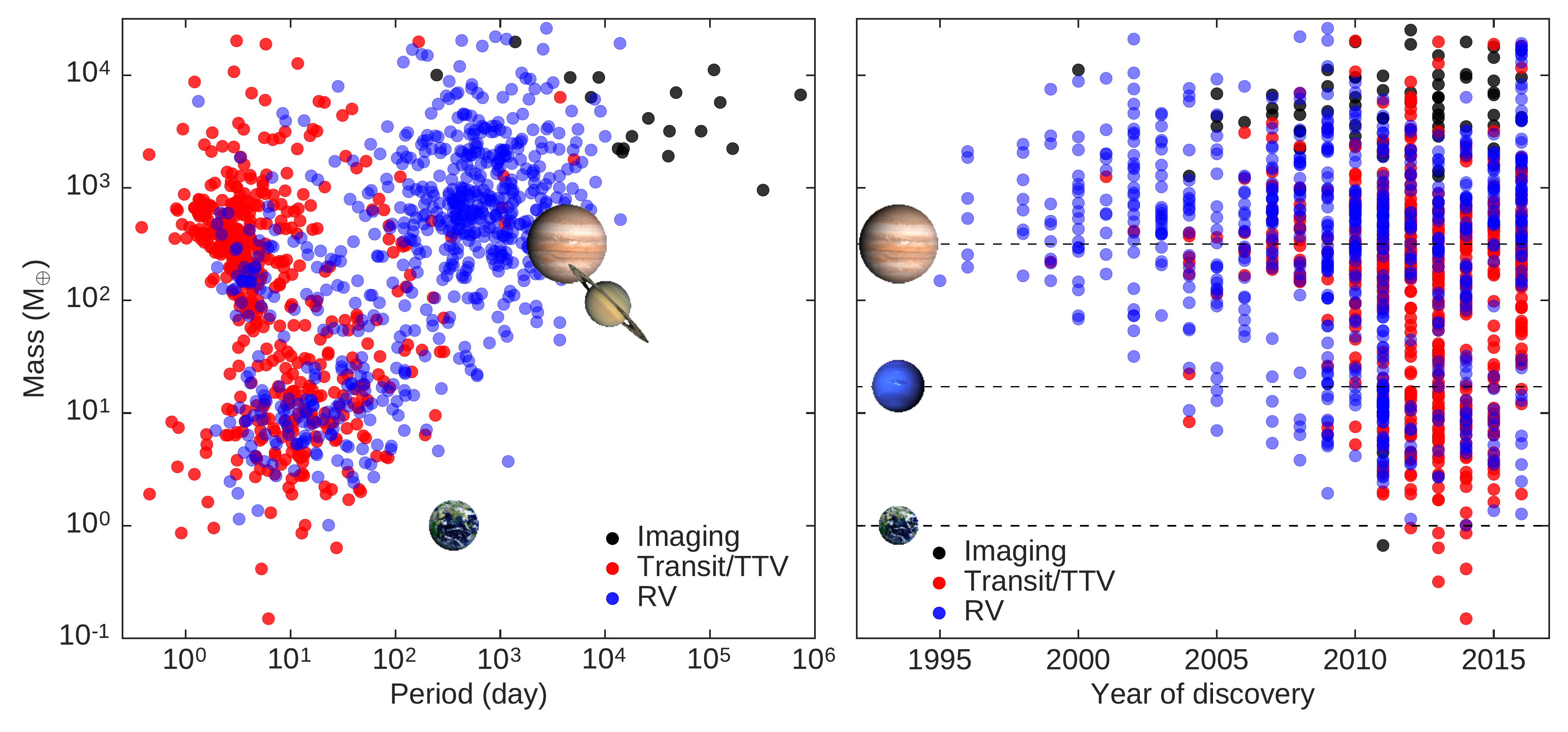}{fig1}
{\textit{Left}: The distribution of discovered planets in the period--mass diagram. \textit{Right}: Mass of the known planets as a function of the 
discovery year. Different symbols represents planets discovered by different detection techniques. Some of the planets of our Solar System are shown for reference.}

The distribution of exoplanets in the period -- mass diagram is shown in Fig.~\ref{fig1}. First, the plot shows that different planet detection techniques
occupy different regions of this diagram. Second, one can see that the detected planets, while being very diverse in their properties (see previous paragraph),
are clustered in three main groups: hot-Jupiters (M$_{p}$ $\sim$1-2M$_{\jupiter}$ and P $\lesssim$ 10 days), hot/warm super-Earths/Neptunes 
(M$_{p}$ $\sim$10M$_{\oplus}$ and P $\lesssim$ 100 days), and gas and ice giants (M$_{p}$ $\sim$1-2M$_{\jupiter}$ and P $\sim$ 1000 days). We note that this
diagram is strongly constrained by the biases and detection limits of different techniques. In particular the detection limits of these techniques and current instrumentation
is responsible for the empty bottom-right triangle of the figure.
However, some of the  observed features, such as the ``period-valley''  \citep[a lack of giant planets with periods between 10-100 days:][]{Udry-03} or the sub-Jovian desert 
\citep[a lack of sub-Jupiter mass planets at orbital periods shorter than 3 days:][]{Szabo-11} are probably physical and give important insights for 
our understanding of exoplanet formation and evolution. For a recent excellent review on the architecture of exoplanetary systems we refer the reader to \citet{Winn-15}.

A logical question now to ask is how do these very different planets form? Currently two main mechanisms are proposed for the formation of exoplanet
that are conceptually different. In the so called core-accretion (CA) model low-mass planets  form 
from the coagulation of very small solid bodies \citep{Pollack-96}. If before the dissipation of the protoplanetary disk, a core of about 5-10 M$_{\oplus}$
is formed then, it can undergo runaway accretion of gas and form a giant planet. In the so called gravitational instability (GI), 
in the gaseous disk (usually massive and cold), localized instabilities collapse into giant planets \citep{Boss-98}. These two models have experienced substantial development and modifications,
and the most recent and advanced ones \citep[e.g.][]{Nayakshin-16, Bitsch-15, Levison-15} include important phenomena such us 
pebble accretion \citep[e.g.][]{Johansen-10} and/or migration in the disk \citep[e.g.][]{Alibert-04}. Planetary population synthesis calculations 
\citep{Ida-04} based both on CA \citep[e.g.][]{Mordasini-09, Hasegawa-13} and GI followed by tidal downsizing \citep[TD -- e.g.][]{Forgan-13, Nayakshin-16} 
reproduce many of the properties of the observed exoplanets. We refer the reader to \citet{Mordasini-15} for a recent review on 
the \textit{Global models of planet formation and evolution}.

\vspace{-0.3cm}

\section{Exoplanets and stellar metallicity}

The correlation between stellar metallicity and the occurrence rate of giant planets is a firmly established fact \citep[e.g.][]{Gonzalez-97, Santos-01},
however, the exact functional form of this dependence is not fully established yet \citep[see left panel of Fig.~\ref{fig2};][]{Mortier-13}.
This observational result got its theoretical support first in CA \citep[e.g.][]{Mordasini-09} and then in the TD 
\citep[][]{Nayakshin-16}\footnote{Note that most of the GI based models do not predict a strong correlation between giant planet frequency and 
metallicity \citep[e.g.][]{Boss-98}.}. Despite the large amount of observational data, it is still not clear if the planet-metallicity correlation
holds for low-mass/small-sized planets \citep[see right panel of Fig.~\ref{fig2}; ][]{Sousa-11, Mayor-11, Wang-15, Buchhave-15, Zhu-16}. This is probably 
because it is hard to detect these light planets (especially at large distances) and it is very difficult to create a comparison sample of stars without low-mass planets.

\vspace{-0.2cm}

\articlefigure[width=0.75\textwidth]{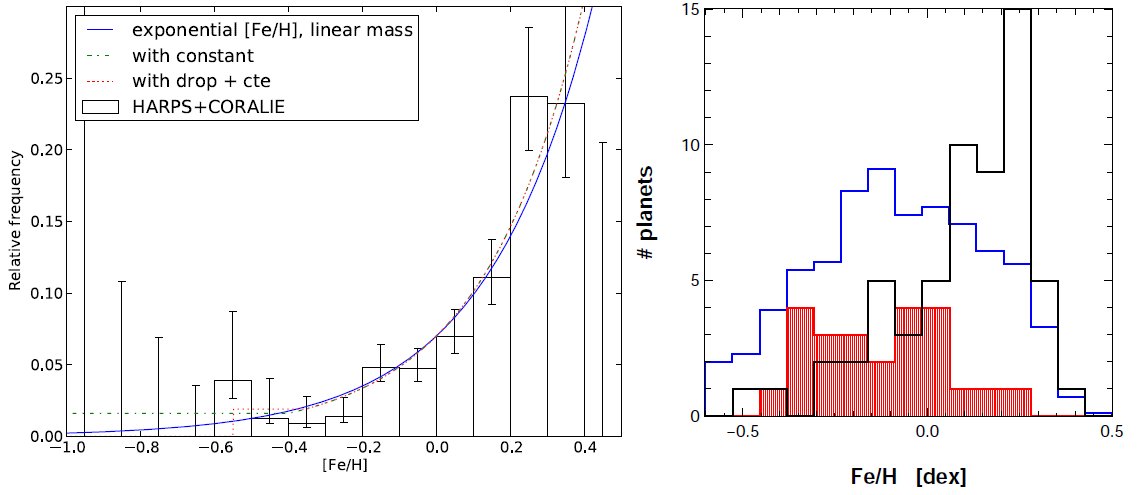}{fig2}
{\textit{Left - \citet{Mortier-13}}:  Frequency of giant planets as a function of metallicity and mass of the HARPS + CORALIE sample. Different functional forms are
shown in different colors . \textit{Right - \citet{Mayor-11}}: The metallicity ([Fe/H]) distribution of stars hosting giant gaseous planets (black), planets 
less massive than 30 M$_{\oplus}$ (red), and for the global combined sample stars (blue). The latter histogram has been divided  by 10 for the sake of visual comparison.}

The importance of metallicity is not only limited to the formation efficiency of planets. Metallicity also determines the maximal mass of the exo-Neptunes \citep{Courcol-16},
the presence or absence of gaseous atmosphere of small-sized planets \citep{Dawson-15}, and the mass of the core (heavy elements) of giant planets \citep[e.g.][]{Miller-11}.

\vspace{-0.4cm}

\articlefigure[width=0.7\textwidth]{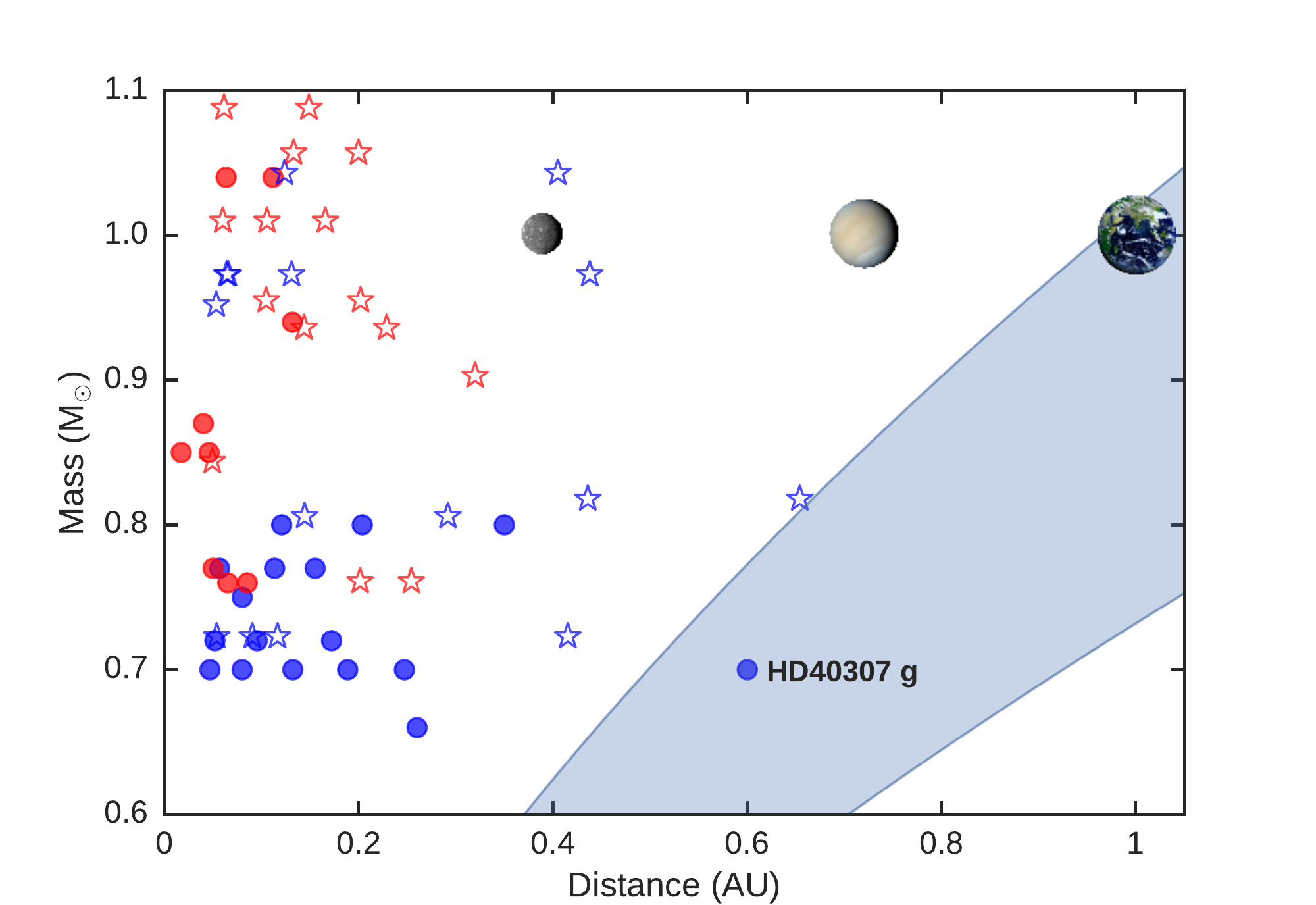}{fig4}
{The orbital semi-major axis of low-mass and small-size planets orbiting FGK dwarf stars. Planets detected by the RV and Transit techniques are 
shown with filled circles and empty star-symbols, respectively. Blue color corresponds to planets orbiting metal-rich stars ([Fe/H] $\geq$ -0.1)
and red color corresponds to planets around metal-poor stars ([Fe/H] $<$ -0.1). The habitable zone \citep[][]{Kopparapu-13} for stars of different masses  is highlighted by
blue shade.} 

It is also interesting to note that the final orbital separation of planets shows a dependence on metallicity of the system \citep{Adibekyan-13, Beauge-13, Mulders-16, Adibekyan-16}.
\citet{Adibekyan-16}, based on the previous results that  low-mass and small-sized planets orbiting around metal-rich stars do not have long orbits \citep{Adibekyan-13},
suggested that planets in the ``habitable zone'' should be preferentially less metallic than our Sun. Fig.~\ref{fig4} shows the orbital distance of low-mass 
(detected with RV) and small-radius (detected by transit method) planets against the mass of their host stars. The plot is based on the data of \citet{Adibekyan-16}
and illustrates their findings. Here we should note that \citep{Mulders-16} observed several \textit{Kepler} planet candidates orbiting their metal-rich stars at 
long periods\footnote{Note that the sample of \citep{Mulders-16} consists of \textit{Kepler} planet candidates and not only confirmed planets.}.
However, they also observed that the planet occurrence rate is two times higher for metal-poor systems when compared to the systems with super-solar metallicities.

\section{Exoplanets and stellar chemistry}

In stellar astrophysics, the iron content is usually used as a proxy for overall metallicity and most of the aforementioned studies followed this trend.
Several works, however, searched for chemical peculiarities of planet hosting stars in terms of abundances of individual elements. While many contradictory 
results can be found in the literature \citep[e.g.][]{Bodaghee-03, Robinson-06, Brugamyer-11, Suarez-16a, Suarez-16b}, 
the enhancement of $\alpha$-elements of iron-poor planet hosts was shown to be robust \citep[][]{Haywood-08, Kang-11, Adibekyan-12a, Adibekyan-12b}.
Interestingly, \citet[][]{Adibekyan-12a} showed that even low-mass/small-radius planets show $\alpha$-enhancement at low-iron regime. 
The right panel of Fig.~\ref{fig3} depicts the $\alpha$-enhancement (here Si abundance is used as a proxy for $\alpha$-elements) of
iron-poor planet hosts for the HARPS sample of \citet{Adibekyan-12c}. The enhancement in $\alpha$-elements relative to iron is typical for the thick disk stars
\citep[e.g.][]{Fuhrmann-98, Adibekyan-13b}.
In fact, the HARPS data suggests that the planet formation frequency is about 5.5 times higher in the thick disk (12.3$\pm$4.1\%) when compared to the 
Galactic thin disk (2.2$\pm$1.3\%) in the metallicity range of -0.7 $<$ [Fe/H] $<$ -0.3 dex.

\citet{Gonzalez-09} recommended to use a so-called refractory index 
``Ref'', which quantifies the mass abundances of refractory elements (Mg, Si and Fe) important for planet formation, rather than [Fe/H]. The importance of
this index increases in the Fe-poor region \citep{Adibekyan-12c, Gonzalez-14} when one compares statistics of planets around the thin disk and thick disk stars.

\vspace{-0.2cm}

\articlefigure[width=0.8\textwidth]{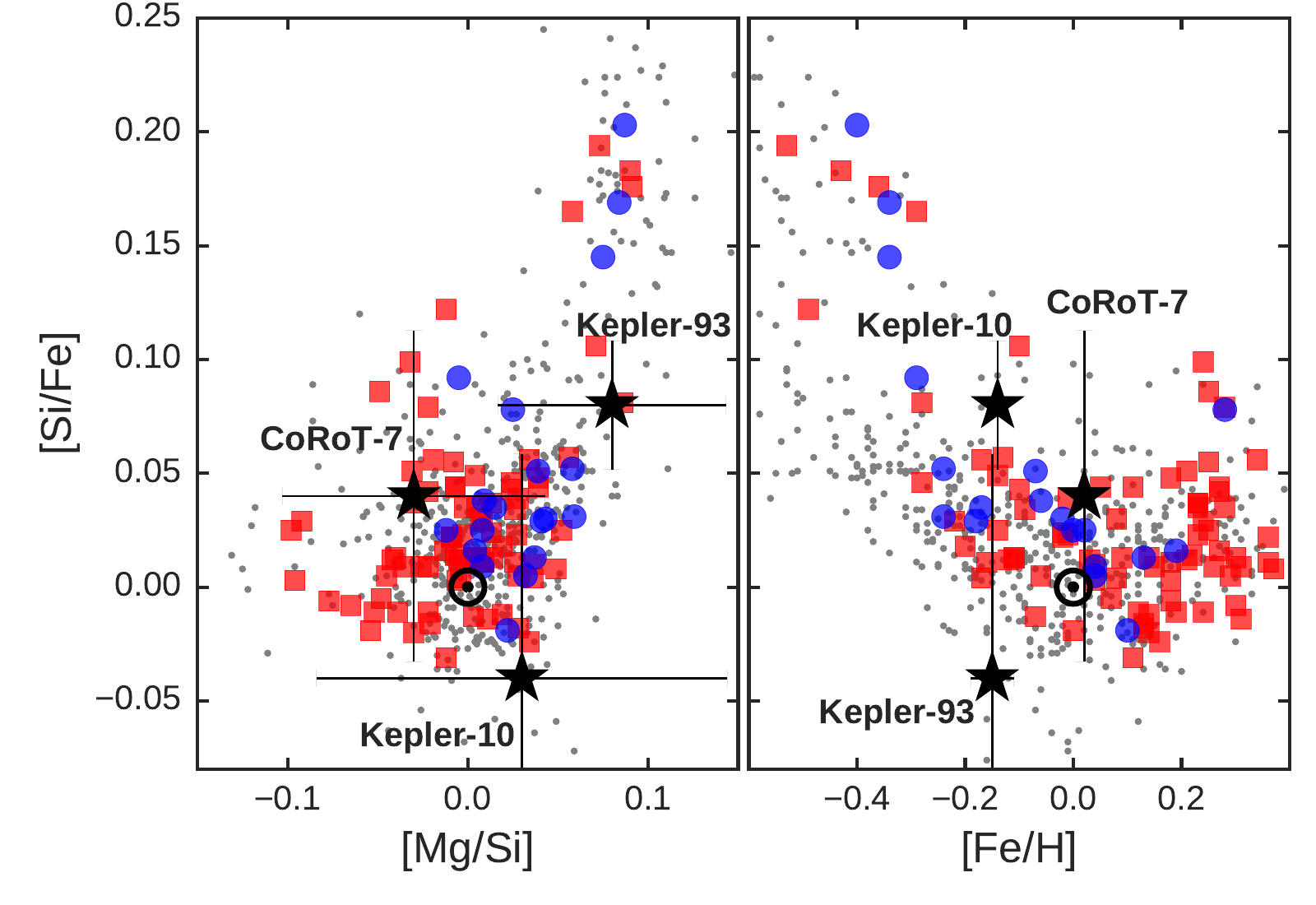}{fig3}
{[Si/Fe] versus [Mg/Si] and [Si/Fe] versus [Fe/H] for stars with and without (gray dots) planets. 
The red squares refer to the Jovian hosts and the blue circles refer to the stars hosting exclusively Neptunians and super-Earths (M $<$ 30 M$_{\oplus}$).
Three stars that are hosting low-mass/small-radius planets with precise radius and mass determinations are presented with a symbol of star \citep{Santos-15}.
The position of the Sun is marked with the modern sun symbol.}

The studies of individual heavy elements and specific elemental ratios  in stars with planets are very important because they are expected to control the structure
and composition of terrestrial planets \citep[e.g.][]{Grasset-09, Bond-10, Thiabaud-14, Dorn-15}. In particular, Mg/Si and Fe/Si
ratios are important to constrain the internal structure of terrestrial planets \citep{Dorn-15}. Recently, \citet{Santos-15} tested these models
on three terrestrial planets (see Fig.~\ref{fig3}) and showed that the iron mass fraction inferred from the mass-radius relationship 
is in good agreement with the iron abundance derived from the host star's photospheric composition.

\section{Exoplanets and Galactic chemical evolution}

As discussed in the previous sections exoplanet formation efficiency, the type of the planets formed and their orbital characteristics depend on metallicity 
and chemical conditions. Putting all these results together one can reach to an interesting conclusion (or perhaps a speculation): i) Planets orbiting their stars 
in the circumstellar habitable zone 
have sub-solar metallicities (Fig.~\ref{fig4}). ii) These iron-poor stars are usually enhanced in $\alpha$-elements (i.e. high Si/Fe ratio) and at the same 
time have high Mg/Si ratio (Fig.~\ref{fig3}). iii) High [Mg/Si] and low [Fe/Si] abundance ratio should produce a planet of a composition and structure that is different than
ours \citep{Dorn-15}. 
Metallicity and abundance of different elements important for planet formation varies with time and location in our Galaxy and in Universe in general.
Several studies during the last decade tried to predict prevalence of terrestrial planets in our Galaxy and in the so called ``Galactic habitable zone''\footnote{The
region within the Galaxy where planets can form around stars and retain liquid water on their surface for a significant amount of time.}
\citep[e.g.][]{Lineweaver-04, Gowanlock-11, Gonzalez-14a, Gobat-16}. Some other studies extended these works to the observable Universe 
\citep[e.g.][]{Behroozi-15, Zackrisson-16}. 
We refer the reader to these interesting works for more information and details about the recent interpretations of evolution of life across space and time.

\section{Conclusion}

Formation efficiency, composition, structure, and even ``habitability'' of planets depend on 
the chemical conditions of the environment they form i.e. time and place in the Galaxy.

\acknowledgements
I thank the science organizing committee of \textit{Non-Stable Universe: Energetic Resources, Activity Phenomena and Evolutionary Processes}  
for the invitation to participate to this interesting conference.
I acknowledge the support from Funda\c{c}\~ao para a Ci\^encia e Tecnologia (FCT) through national funds
and from FEDER through COMPETE2020 by the following grants UID/FIS/04434/2013 \& POCI-01-0145-FEDER-007672, 
PTDC/FIS-AST/7073/2014 \& POCI-01-0145-FEDER-016880 and PTDC/FIS-AST/1526/2014 \& POCI-01-0145-FEDER-016886. 
I also acknowledges the support from FCT through Investigador FCT contracts of reference IF/00650/2015.
Finally, I would like to thank Pedro Figueira for his interesting comments and discussion.

\bibliography{references.bib}

\end{document}